\documentstyle[ibvs]{article}
\begin{document}


\def\ergs{erg s$^{-1}$}
\def\ergcm2sa{erg cm$^{-2}$ s$^{-1}$ \ang$^{-1}$}
\def\dsct{$\delta$~Sct}
\def\ang{\AA}

\IBVShead{4264}{? November 1995}

\title{New Photometry of the Hyades $\delta$ Scuti Star V777 Tau (71 Tau)}

71 Tauri (= BS 1394 = vB 141 = V777 Tau) is a rapidly rotating, F0 IV--V
member of the Hyades cluster.  It is the second brightest
X-ray source in the cluster, after V471 Tauri, and an
intensely bright coronal EUV source (Stern {\em et al.} 1995).
It is a lunar occultation binary, but the secondary star (estimated by
Peterson {\em et al.} 1981 to be a G4 V star in
possibly a 53 day orbit) has never been seen directly.  As
pointed out by Stern {\em et al.}, the tremendous X-ray
luminosity of 71 Tau, $L_x = 10^{30}$ \ergs, is not the
result of a flare.  It is not easily explained as the coronal
emission of an individual star, or even as the combination
of emission from several stars, considering that most F--M
Hyades stars are detected at levels of only $L_x \sim$
few~$\times 10^{29}$ \ergs (or in some cases, much less).
The X-ray properties of 71 Tau are more in line with those of a very
much younger star, but would still be considered remarkable even for a
coronally active star in the $\approx$80 Myr old Pleiades cluster
(Stauffer {\em et al.} 1994).  71 Tau has been observed a number of
times by IUE and has been shown to be variable by as much as
30 \% at ultraviolet wavelengths near 1700--2000\ang\
(Simon and Landsman, in preparation).

Horan (1979) discovered that 71 Tau was a $\delta$ Scuti star with an
amplitude of 0.01 to 0.02 mag.  To our knowledge no other optical photometry
of 71 Tau has ever been published.  Horan gives a principal period of 3.9 hours
($f$ = 6.15 d$^{-1}$) and suggests that 71 Tau may be excited in more than one
mode.  However, his conclusions are based on only 38 points obtained on two
nights which were two months apart.  From Horan's Figure 5 it seems much
more likely that the  principal period is near 4.4 hours if the decline
in brightness occurs at the same rate at the increase of brightness.

We have obtained new photometry of 71 Tau, observed differentially with
respect to BS 1422, amounting to 58 points on 5 nights
during a 7 night run with the 0.6-m telescope at Mauna Kea.
Observations of BS 1432 (the check star) vs. BS 1422 indicated that both
were constant, so any variations in the differential magnitude of 71 Tau
vs. BS 1422 may be attributed to 71 Tau.
The individual data points can be obtained by requesting file
307E from IAU Commission 27, Archives of Unpublished Photometry (see Breger
{\em et al.} 1990).

Figure 1 shows a power spectrum of the photometry of 71 Tau vs. BS 1422,
using the Lomb-Scargle algorithm as presented by Press and Teukolsky (1988).
Clearly, many aliases are present.  On the basis of Horan's Figure 5
and folded plots of our data we believe the frequency near 5.5 cycles
per day is more likely to be the true principal frequency, not its
one-day alias at 6.5 d$^{-1}$.  (Note that the frequency corresponding to
Horan's period is between these two values.)
Having settled on a principal period
of 0.1823 day, we subtracted a properly phased least-squares sinusoid
with that period from the data to see if other frequencies are present.
That power spectrum is shown in Figure 2.  We note that if $f_{1} \approx$
6.5, essentially the same power spectrum of residuals results.

One might assume that one of the peaks in Figure 1 represents pulsation in
the fundamental radial mode, while one of the peaks in Figure 2 represents
the first overtone radial mode.  However, the period ratio of first overtone
to fundamental should be close to the observed value of 0.773 (Breger
1993) -- essentially equal to the theoretical value of 0.772 (Guzik and Bradley
1995).  The closest we can come is $1/8.63 \div 1/6.49 \approx 0.752$.  But
given the small number of data points, we do not feel that this is the best
method for choosing which peaks in the power spectrum are true and which are
aliases.

Using an epoch of HJD 245\,0000, we believe the following is the best
two-frequency fit to the data: $f_{1}$ = 5.485 d$^{-1}$, amplitude =
6.0 $\pm$ 0.7 mmag, phase = 0.171 $\pm$ 0.016; $f_{2}$ = 7.637 d$^{-1}$,
amplitude = 3.4 $\pm$ 0.7 mmag, phase = --0.439 $\pm$ 0.030.  Thus, we find
a principal period of 0.1823 days (4.38 hours), and a secondary period of
0.1309 days (3.14 hours). The ratio of the latter to the former is 0.718.
Given the amplitudes of the two frequencies, we can account for a range in
brightness up to 0.02 mag in $V$, and we can also account for the differing
amplitudes observed by Horan on the two nights he measured the star.

Figure 3 shows our data folded by the principal period after the data have
been prewhitened by the secondary frequency.

A definitive solution to the photometric behavior of 71 Tau could
only be obtained from a more extensive data set than ours.   In
light of the unusual properties of this star at X-ray and ultraviolet
wavelengths, such further study seems amply warranted.

\begin{center}
{\bf Acknowledgments}
\end{center}
KK thanks the University of Hawaii for telescope time on the 0.6-m telescope
and thanks the Joint Astronomy Centre for observing support.  MR's observing
expenses were paid for by a University of Hawaii at Hilo fund endowed by
William Albrecht. Luis Balona kindly provided the program for obtaining
least-squares Fourier fits to the data.

\references

Breger, M., 1993, in IAU Colloq. 139, ed. J. Nemec and J. M. Matthews,
  Cambridge Univ. Press, 135

Breger, M., Jaschek, C., and DuBois, P., 1990, IBVS No. 3422

Guzik, J. A., and Bradley, P. A., 1995, in {\em Astrophysical Applications
  of Stellar Pulsation}, IAU Colloq. 155, ed. R. S. Stobie and P. A.
  Whitelock, ASP Conf. Series, vol. 83, 92

Horan, S., 1979, AJ, 84, 1770

Peterson, D. M., {\em et al.}, 1981, AJ, 86, 1090

Press, W. H., and Teukolsky, S. A., 1988, Comput. Phys., 2, No. 6 (Nov/Dec),
  77

Stauffer, J. R., {\em et al.}, 1994, ApJS, 91, 625

Stern, R. A., {\em et al.}, 1995, ApJ, 427, 808

\vspace {5 mm}

KEVIN KRISCIUNAS, Joint Astronomy Centre, 660 N. A'oh\={o}k\={u} Place,
University Park, Hilo, Hawaii 96720 USA \\
email: kevin@jach.hawaii.edu\\

THEODORE SIMON, University of Hawaii, Institute for Astronomy,
2680 Woodlawn Drive, Honolulu, Hawaii 96822 USA \\
email: simon@hubble.ifa.hawaii.edu\\

RICHARD A. CROWE, MAVOURNEEN ROBERTS, Department of Physics and Astronomy,
University of Hawaii at Hilo, 200 West Kawili Street, Hilo, Hawaii 96720 USA\\
email: rcrowe@maxwell.uhh.hawaii.edu; morning@einstein.uhh.hawaii.edu\\

\vspace {10 mm}

\begin{center}
{\bf Figure Captions}
\end {center}

\vspace {5 mm}

Figure 1 -- Power spectrum of $V$-band differential photometry of 71 Tau
vs. BS 1422.  The frequency corresponding to Horan's (1979) period
is 6.15 d$^{-1}$.

\vspace {5 mm}

Figure 2 -- Power spectrum of the data after prewhitening the data by a
least-squares sinusoid with $f_{1}$ = 5.485 d$^{-1}$.

\vspace {5 mm}

Figure 3 -- Folded light curve of 71 Tau vs. BS 1422 $V$-band data, after
prewhitening by the secondary sinusoid with $f_{2}$ = 7.637 d$^{-1}$.

\end{document}